%% file: final.tex
\documentclass[twoside]{article}

\usepackage{graphicx}
\usepackage{algorithm}
\usepackage{algorithmic}
\usepackage{subfigure}
\usepackage{hyperref}
\graphicspath{{figure/}}
\usepackage{amsthm}
\newtheorem{thm}{\bf Theorem}[section]
\newtheorem{Corollary}[thm]{\bf Corollary}
\usepackage[accepted]{aistats2020}
\usepackage[dvipsnames]{xcolor}
\begin{document}

\twocolumn[

\aistatstitle{An approximate {KLD} based experimental design for models with intractable likelihoods}

\aistatsauthor{ Ziqiao Ao \And Jinglai Li }

\aistatsaddress{ School of Mathematical Sciences\\ Shanghai Jiao Tong University \And
School of Mathematics\\  University of Birmingham} 
]

\begin{abstract}
Data collection is a critical step in statistical inference and data science,
and the goal of statistical experimental design (ED) is to find the data collection setup
that can provide most information for the inference. 
In this work we consider a special type of ED problems where the likelihoods are not available in 
a closed form. 
In this case, the popular information-theoretic Kullback-Leibler divergence (KLD) based design criterion
can not be used directly, as it requires to evaluate the likelihood function. 
To address the issue, we derive a new utility function,
which is a lower bound of the original KLD utility. 
This lower bound is expressed in terms of the summation of two or more entropies in the data space, 
and thus can be evaluated efficiently via entropy estimation methods.
We provide several numerical examples to demonstrate the performance of the proposed method.   
\end{abstract}

\section{Introduction}

Collecting data is a critical step in statistical inference.  
In practice the data collection procedure may require considerable resources financially and in kind, 
and thus how to effectively allocate the limited resources in the data collection exercise becomes a question of essential importance. 
Statistical experimental design (ED) seeks to address the problem by
developing systematical rules for allocating the resources in the data collection exercise~\cite{ryan2016review}. 

In practice, the ED problems are often formulated as an optimization program, i.e., to optimize the experimental conditions 
with respect to certain design criterion~\cite{fedorov2013theory}.  These design criteria are referred to as the utility functions, as they are meant to measure how useful the experiment outcomes, i.e., the collected data, are. 
It is easy to see that choosing a sensible utility functions is the key in an ED problem, and in the Bayesian inference setting, the Kullback-Leibler divergence  between 
the prior and the posterior distributions is arguably the most popular choice for the utility function, for that it has  certain theoretical merits \cite{paninski2005asymptotic}.  
Intuitively speaking the KLD based criterion seeks to find the experimental conditions that maximize  the information gain expected from conducting experiments. 

As will be shown later, the use of the KLD utility requires to evaluate the likelihood function. In many real-world inference problems, e.g., the biological process models~\cite{cook2008optimal,ryan2016optimal} the likelihood functions may be intractable, .i.e.,
not available in a closed form, and as a result it is not possible to apply the KLD  based ED directly to such problems. 
Considerable efforts have been devoted to developing  likelihood-free ED method for such problems. 
Some commonly used methods rely on the ability to approximate the likelihood,  or related functions, and these methods include \cite{overstall2018bayesian,dehideniya2019synthetic,pmlr-v89-kleinegesse19a}. 
{Another} popular class of likelihood free methods are based on the Approximate Bayesian Computation (ABC) \cite{beaumont2002approximate} and the posterior covariance based design utility. 
Works in this class include \cite{drovandi2013bayesian, hainy2016likelihood,dehideniya2018optimal}, and so on.
A major limitation of the posterior covariance based design utility is that it {may not work} when the posterior distributions deviate significantly from Gaussian. 

The main goal of this work is to provide a alternative method for ED problems with intractable likelihoods, which can be used regardless whether the posterior is close to Gaussian. 
Specifically, we propose a utility  that is {essentially} a lower bound approximation of the KLD utility (namely, it is equivalent to maximizing a lower bound of the original KLD utility),
and we then present a numerical scheme based on entropy estimation to evaluate this utility function  without querying the likelihoods. 
Finally we provide numerical examples to demonstrate the performance of the proposed ED method. 

 The reminder of the paper is organized as follows. Section \ref{sec:bed} reviews 
the set up of the Bayesian experimental design (BED), the KLD based design, and the methods to address intractable likelihoods.
Section \ref{sec:method} provides our new design utility as well as an entropy estimation based numerical scheme to compute it. 
Finally Section \ref{sec:examples} provides several examples to illustrate the performance of the proposed method. 

\section{Background}
\subsection{Bayesian Experimental Design}\label{sec:bed}

The Bayesian inference problems with controllable design parameters can be described as follows. 
Let $y\in \mathcal{Y}$ be the data that are observed and $\theta\in\Theta$ be the unknown parameters that we want to estimate.
The relation between $\theta$ and $y$ is characterized by the likelihood function $p(y|\theta,d)$,
where $d$ is the design parameters representing the experimental conditions that can be controlled by the users. 
The parameter of interest $\theta$ can be inferred from observed data $y$ by computing the posterior distribution via Bayes’ formula
$$p(\theta|y,d)=\frac{p(\theta|d)p(y|\theta,d)}{p(y|d)},$$
where $p(\theta|d)$ is the prior distribution of $\theta$ and $p(y|d)$ is the evidence of model. 

As is mentioned earlier, due to time and monetary limitations, one usually can only afford to conduct a fixed number of experiments and collect their outcomes, i.e., the data. 
To this end, the goal of the experimental design is to identify the experimental conditions (represented by the design parameter $d$) which is the most useful for the inference task. 
Mathematically the ``usefulness'' of the experimental condition can be defined by a utility function:
$u(d,y,\theta)$, which measures the worth of the experiment with true model parameter $\theta$ that applies design $d$ and yields an observation $y$.
Then one can determine the value of the design parameter $d$  by maximizing the expected utility:
\begin{equation}
	\max_{d\in \mathcal{D}}U(d)=\int_{\mathcal{Y}}\int_{\Theta}u(d,y,\theta)p(\theta,y|d)d\theta dy, \label{e:Ud}
\end{equation}
where $\mathcal{D}$ is the state space of $d$. 
As is mentioned earlier, the KLD between the posterior and the prior distributions,
 namely,
\begin{equation}
	u(d,y):=\int_{\theta}p(\theta|y,d)log\left[\frac{p(\theta|y,d)}{p(\theta|d)}\right]d\theta,
\end{equation}
is a popular choice of the utility function for the Bayesian inference problems.
One main challenge here is to solve the optimization problem in Eq.~\eqref{e:Ud}, and to do so we need to evaluate the objective function $U(d)$. 
In general the function $U(d)$ does not admit an analytical expression and needs to be evaluated numerically. 
First we write the function $U(d)$ as, 
\begin{equation}
		U(d)
		=\int_{\mathcal{Y}}\int_{\Theta}\{ \log[p(y|\theta,d)] - \log[p(y|d) \}p(y,\theta|d)dyd\theta, \label{e:kleval}
\end{equation}
%where $H(p(y|d)$ is the entropy of distribution $p(y|d)$:
%\begin{equation}
%H[p(y|d)] = \int \log[p(y|d)]p(y|d) dy.
%\end{equation}
It follows from Eq.~\eqref{e:kleval} that $U(d)$ {can be estimated} via a MC simulation: 
\begin{equation}
	{U}(d)\approx \frac{1}{n}\sum_{i=1}^{n}\{log(p(y_i|\theta_i,d))-log(p(y_i|d))\} \label{e:Umc}
\end{equation}
where $\{(\theta_i,y_i)\}_{i=1}^n$ are drawn from $p(y,\theta |d)$. 
Note however here that in usual Bayesian inference problems $p(y|d)$ is not known in advance and has to be estimated via MC as well:
\begin{equation}
	p(y_i|d)=\int_{\Theta}p(y_i|\theta,d)p(\theta|d)d\theta\approx\frac{1}{n'}\sum_{j=1}^{n'}p({y_i}|\theta_{i,j},d), \label{e:lkmc}
\end{equation}
where $\{\theta_{i,j}\}_{j=1}^{n'}$ are  samples from the prior $p(\theta|d)$. 
By combining Eq.~\eqref{e:Umc} and Eq.~\eqref{e:lkmc}, we obtain a nested MC estimator of $U(d)$:
\begin{equation}
	{U}(d)\approx\frac{1}{n}\sum_{i=1}^{n}\{\log(p(y_i|\theta_i,d))-\log(\frac{1}{n'}\sum_{j=1}^{n'}p(y_i|\theta_{i,j},d))\}.
\end{equation} 
Once being able to evaluate the objective function $U(d)$, one  can solve the optimization problem~\eqref{e:Ud} by some derivative free algorithm, see e.g., \cite{huan2013simulation}.

\subsection{BED with intractable likelihoods}
We have provided a brief introduction to the BED problem in Section~\ref{sec:bed}. 
It is obvious that the procedure described in Section~\ref{sec:bed} requires the ability to evaluate the likelihood function $p(y|\theta,d)$.
 which, in many complex practical problem,  may not be possible. 
More specifically, we shall assume that, in these problems, one can draw samples from the likelihood function, but 
can not evaluate the likelihood function directly. 
To this end, several methods have been proposed to conduct the likelihood-free experimental design, and most of them are 
based on the ABC method for sampling the posterior. Though these methods are different technically, the main ideas behind them are quite similar: 
one should seek the design parameter that yields the posterior distribution with the minimal ``uncertainty'',
which is measured by certain quantities  associated with the posterior covariance.  
The posterior distribution here is computed by the ABC method.

We give a brief description of one of this type methods.  
For example, one can chose the utility function to be the inverse of the determinant of the posterior covariance matrix:
\begin{equation}
u(d,y)=\frac{1}{\det(\mathrm{cov}(\theta|y,d))},
\end{equation}
which is numerically evaluated by, 
\begin{equation}
u(d,y)\approx \frac{1}{\det(\hat{C}(\theta_1,...,\theta_n))},
\end{equation}
where $\{\theta_i\}_{i=1}^n$ is an ensemble drawn from the posterior distribution $p(\theta|y,d)$
and $\hat{C}$ is the sample covariance of it. 
Obviously to evaluate such a $u(d,y)$ one need to draw samples from the posterior distribution $p(\theta|y,d)$.
Using the ABC methods \cite{beaumont2002approximate}, one can draw samples from the posterior distribution 
without evaluating the likelihood. 
This way, one can evaluate the utility function $u(d,y)$ for any given pair of $(d,y)$. 
Finally, the expected utility $U(d)$ is then computed via MC:
\begin{equation}
U(d)=\frac{1}{n}\sum_{i=1}^{n}{u}(d,y_i),
\end{equation}
where $\{y_i\}_{i=1}^n$ are drawn from the joint distribution $p(y,\theta)$.  
Following \cite{drovandi2013bayesian}, we refer to this approach as the \emph{D-posterior precision} method.  
Other utility function associated with the posterior covariance can also be used, 
e.g. the inverse of the trace of the posterior covariance.
By using the ABC method and utility function based on the posterior covariance,
one can conduct the experimental design without evaluating the likelihood function. 
However, a major limitation of this type of methods is that 
the use of such utility may be inappropriate if the posterior distributions
deviate significantly from Gaussian (e.g., multimodal distributions).
 as in that case the uncertainty of the posterior {may} 
not be well quantified by its variance.

The aforementioned KLD utility can alleviate this limitation and be used for any posterior distribution.
Unfortunately, as is pointed out at the beginning of the section, computing the KLD utility is extremely challenging when the likelihood is intractable. 
Several works such as \cite{overstall2018bayesian}, \cite{pmlr-v89-kleinegesse19a} propose to address the issue by approximating or numerically estimating the likelihood or related functions, and 
accurate estimation or approximation of these functions may be either computationally intensive 
or assumes the likelihood to be of certain specific form. 
%In particular,  \cite{overstall2018bayesian} approximates the likelihood function using the Gaussian process model,
%and \cite{pmlr-v89-kleinegesse19a} estimates the ratio function between the posterior and the prior;
 %\cite{terejanu2012bayesian} converts the expected KLD utility into 
%into the mutual information between $y$ and $\theta$ and estimates it with a kernel density estimator proposed by \cite{kraskov2004estimating}. 
%The first two methods  require to approximate the likelihood or a related function, while \cite{terejanu2012bayesian} nees
An alternative solution is to evaluate the expected KLD utility by {entropy estimation}. 
To this end, \cite{terejanu2012bayesian} evaluates the expected utility by estimating the mutual information in the joint 
space of $\theta$ and $y$, which is impractical when the unknown $\theta$ is high dimensional.   
In this work, we adopt the entropy estimation framework, and in the next section we propose a new 
utility function which only requires to estimate a small number of entropies in the output space.

\section{The entropy estimation based BED method}\label{sec:method}

In this section we introduce the approximate KLD utility 
and its numerical implementation. 
\subsection{A lower bound approximation of the expected KLD utility}

By some elementary calculus we can rewrite Eq.~\eqref{e:kleval} as, 
\begin{equation}
	%\begin{aligned}
		U(d)=-E_\theta[H(p(y|\theta,d))]+H(p(y|d)), \label{e:ud2}
	%\end{aligned}
\end{equation}
where $H(p(\cdot))$ is defined as the entropy of distribution $p(\cdot)$:
\begin{equation}
H[p(y)] = - \int \log[p(y)]p(y) dy.
\end{equation}
A very intuitive idea to evaluate $U(d)$ is to  apply an MC estimation to the first term in Eq.~\eqref{e:ud2}:
\begin{equation}
U(d) \approx -\sum_{i=1}^n H(p(y|\theta_i,d)) +H(p(y|d)),
\end{equation} 
where $\theta^1...\theta^n$ are drawn from the prior $p(\theta)$. 
One then can use some entropy estimation approach to compute $H(p(y|\theta,d))$ and $H(p(y))$.
%which means that we {\color{blue}can $p(y|\theta,d)$}. 
Even though the procedure described above can evaluate $U(d)$ with intractable likelihood function,
it may not be practical as it requires to perform entropy estimation $n+1$ times. 
In reality, the entropy estimation is computationally intensive, and $n$, the number of  MC samples are usually large (say, $10^4$ or larger), which may render
 the total computational cost  prohibitive.
 Given that we want to maximize $U(d)$, an intuitive solution  is to  construct a lower bound approximation to $U(d)$, 
 and maximize {this} lower bound  instead. 
 Our construction of the lower bound is based on Theorem~\ref{thm1}.

\begin{thm}\label{thm1}
	Suppose that  $\theta$ is a random variable defined on state space $\Theta$, with probability density $p(\theta)$. For any given $\theta\in\Theta$, let $y$ and $y'$ be two random variables that are independent conditional on $\theta$, and 
	both follow the same distribution $p(y|\theta)$. Now define $z=y-y'$, and we then have,  
	$$E_\theta[H(p(y|\theta))] \leq H(E_{\theta}[p(z|\theta)])-\frac{dim(y)}{2}\log2,$$
 where $dim(y)$ is the dimensionality of $y$.
	%Moreover, equality holds if and only if $X|\theta$s are Gaussian with the same probability density for different $\theta$s.
\end{thm} 

%Using the well known entropy power inequality, we have 
%$$exp(2H(X-X')/p)\geq exp(2H(X)/p)+exp(2H(-X')/p)=2exp(2H(X)/p)$$
%where $X$ and $X'$ are p-dimensional independent variables from the same distribution. Then we have
%$$2H(X)/p+log2\leq 2H(X-X')/p$$
%$$H(X)\leq H(X-X')-\frac{p}{2}log2$$
%Finally, via this inequality as well as the concavity of entropy, we can get a lower bound of the minus conditional entropy term of eq. (4). 
%\begin{equation}
%	\begin{aligned}
%		&-\int_{\Theta}p(\theta)H(p(y|\theta)) d\theta\\
%		&\geq -\int_{\Theta}p(\theta)H(p(y-y'|\theta)) d\theta+\frac{p}{2}log2\\
%		&\geq-H(E_{\theta}[p(y-y'|\theta)])+\frac{p}{2}log2\\
%		&=-H(E_{\theta}[p(z|\theta)])+\frac{p}{2}log2\\
%	\end{aligned}
%\end{equation}
%where $y$ and $y'$ are independent observations generated from the same parameter and $z$ denotes the difference between them. 
The proof of the Theorem~\ref{thm1} is based on the entropy power inequality~\cite{cover2012elements} and entropy's concavity property~\cite{cover2012elements},
and a complete proof is provided in the Supplementary Material.
Applying Theorem~\ref{thm1} to Eq.~\eqref{e:ud2} one obtains  a lower bound of $U(d)$:
\begin{equation}
	U_L(d) = -H(E_{\theta}[p(z|\theta,d)])+\frac{dim(y)}{2}\log2+H(p(y|d))\label{e:ul1}
\end{equation}
Note here that to evaluate this approximation one only need to perform entropy estimation twice. 
Moreover, this lower bound approximation can be further refined,
 based on the following corollary: 

\begin{Corollary}\label{thm2}
	Suppose $p(\theta)$, $p(y|\theta)$, and $p(z|\theta)$ are defined as is in Theorem~\ref{thm1}, and $p(\theta)$  admits the form of,  
	$$p(\theta)=\sum_{l=1}^{L}\omega_l f_l(\theta),$$
	where $\omega_l\geq0$ for $l=1...L$, $\sum_{l=1}^{L}\omega_l=1$, and $f_l(\theta)$ are density functions.  
	Then 
	\begin{align*}
	E_\theta[H(p(y|\theta))] &\leq \sum_{l=1}^{L} \omega_l H(E_{\theta\sim f_l}[p(z|\theta)])-\frac{dim(y)}{2}\log2\notag\\
	&\leq H(E_{\theta}[p(z|\theta)])-\frac{dim(y)}{2}\log2.
	\end{align*}
	%Moreover, equality holds if and only if $X|\theta$s are Gaussian with the same probability density for different $\theta$s.
\end{Corollary} 
The proof of the Corollary is also provided in the Supplementary Material. 
Now we discuss how to design a lower bound based on Corollary~\ref{thm2}. 
First we divide  the observation space $Y$ into a fixed number of disjoined partitions: $\{Y_l\}_{l=1}^L$, 
such that $\cup_{l=1}^L Y_l= Y$ and $Y_l\cap Y_{l'}=\emptyset$ if $l\neq l'$. 
It then follows that the prior distribution can be written as 
\begin{equation}
p(\theta)=\sum_{l=1}^{L}\omega_l f_l(\theta), \label{e:mixture}
\end{equation}
where $f_l(\theta)=p(\theta|y\in Y_l)$ and $w_l=p(y\in Y_l)$ for $l=1...L$. 
In turn we obtain another lower bound approximation of $U(d)$: 
\begin{multline}
	U_{L}(d)=-\sum_{l=1}^{L} \omega_l H(E_{\theta\sim f_l}[p(z|\theta)])\\+\frac{dim(y)}{2}log2+H(p(y)), \label{e:ul2}
\end{multline}
which is tighter than that in Eq.~\eqref{e:ul1}.
In what follows we shall refer to  $U_L(d)$ in Eq.~\eqref{e:ul2} as the expected 
lower-bound (LB)-KLD utility with partition and $U_L(d)$ in Eq.~\eqref{e:ul1} as the expected LB-KLD utility without partition. 
In our method, we choose to use LB-KLD with partition. 
Next we shall discuss how to numerically evaluate this new expected utility function.  

%Now, let us consider how this partition impacts on above bounds.  
%
%Reconsidering the first term of (4), we change the prior from $p(\theta)$ to $\sum_{l=1}^{L}\omega_l f_l(\theta)$,
%\begin{equation}
%	\begin{aligned}
%		&-\int_{\Theta}p(\theta)H(p(y|\theta)) d\theta\\
%		&=-\sum_{l=1}^{L} \omega_l \int_{\Theta} f_l(\theta) H(p(y|\theta)) d\theta\\
%		&\geq -\sum_{l=1}^{L} \omega_l H(E_{\theta\sim f_l}[p(z|\theta)])+\frac{p}{2}log2
%	\end{aligned}
%\end{equation}
%The inequality is due to the use of ineq. (7).

\subsection{Evaluating the lower bound approximation}
First note that the proposed lower bound approximation $U_L(d)$ in Eq.~\eqref{e:ul2} is actually the sum of several entropies. Thus
an effective entropy estimation method is required to compute this approximation. 
In this work we adopt the  Nearest Neighbor (NN) based entropy estimator  provided in \cite{kraskov2004estimating},
while noting that other choices are also available \cite{beirlant1997nonparametric}. 
Using the method, we can estimate the entropy from a given set of samples. 
%Next, using this approach, we will estimate the lower bound only based on samples from observation space.

We summarize the procedure of estimating the lower bound approximation $U_L(d)$ in Algorithm~\ref{alg:ldkld}.
In addition to the design parameter value $d$, the algorithm {users need to} specify two algorithm parameters: the number of samples generated $n$, and the number of partitions of the state space $L$.
In this algorithm the function $\mbox{EntEst}[\cdot]$ takes a set of samples as its input and 
outputs an entropy of these samples estimated using the method in \cite{kraskov2004estimating}.
A key step in the algorithm is to decompose the prior distribution $p(\theta|d)$ into a number of mixture components as is in Eq.~\eqref{e:mixture}.
%which in the numerical implementation is essentially to cluster a set of sample $\{\theta_i\}_{i=1}^n$  drawn from $\pi(\theta|d)$ into a fixed number of groups. 
As is mentioned earlier, the mixture representation of $p(\theta|d)$ is constructed by  partitioning  the output space $\mathcal{Y}$,
and numerically the partition is achieved by clustering the samples in the output space into a fixed number of groups. 
Specifically, we  use a constrained k-means method \cite{berkhin2006survey} to cluster the samples of into $L$ groups subject to the constraint that the size of any group is larger than a fixed value $n_{\min}$.  
Once the clusters of the output samples are determined, the mixture of $p(y|\theta)$ is obtained. 
As one can see that, in the algorithm, we do not need an explicit form of the mixture, and rather we only need the clustering of the samples drawn from the prior distribution. 

\begin{algorithm}[h]
 \caption{LB-KLD Estimator} \label{alg:ldkld}
 \hspace*{0.02in} {\bf Input:} 
 $d$, $n$, $L$\\
 \hspace*{0.02in} {\bf Output:} 
 $\hat{U}_L$
 
 \begin{algorithmic}[1]
  \FOR{$i=1$ to $n$}
  \STATE Generate $\theta_i \sim p(\theta|d)$ and $y_i^*\sim p(y|\theta_i,d)$
  \ENDFOR
  \STATE $\hat{H}^*= \mbox{EntEst}[\{y^*_i\}_{i=1}^n]$
  \STATE $[\widetilde{\Theta},N] = \mbox{Partition}[\{(\theta_i,y^*_i)\}_{i=1}^n,L]$
  %Classify $\{\theta_i\}_{i=1}^N$ into several clusters $\Theta_1,...,\Theta_L$ according to $\{y_i^*\}_{i=1}^N$, where $\Theta_l=\{\theta_{l_j},j=1,...,N_l\}$ and $N=\sum_{l=1}^{L}N_l$
  %\STATE The weights $w_l=\frac{N_l}{N}$
  \FOR{$l=1$ to $L$}
  %\STATE Initial $y,y'$ and $z$
\STATE  $\omega_l=\frac{N(l)}{n}$
  \FOR{$j=1$ to $j=N(l)$}
  \STATE Generate $y_j,y_j' \sim p(y|{x_j}\sim \Theta_l,d)$ //$x_j$ represents the $j$th element of $\Theta_l$
  \STATE Compute $z_j=y_j-y_j'$
  \ENDFOR
  \STATE $\hat{H}_l= \mbox{EntEst}[\{z_j\}_{j=1}^{N(l)}]$
  \ENDFOR
  \STATE $\hat{U}_L=-\sum_{l=1}^{L} \omega_l \hat{H}_l+\frac{dim(y)}{2}\log2+\hat{H}^*$
  \RETURN $\hat{U}_L$
 \end{algorithmic}
\end{algorithm}

\begin{algorithm}[h]
	\caption{ $[\Theta,N] = \mbox{Partition}[\{(\theta_i,y^*_i)\}_{i=1}^n,L]$} 
	\hspace*{0.02in} {User-specified parameters: $n_{\min}$}  \\
	% $L_{\max}$, $\delta$\\
	\begin{algorithmic}[1]
		%\STATE Initiate $L=1$;
		%\STATE Initiate $w=n$; 
		%\WHILE{$min(N)\geq\delta n \& L\leq L_{\max}$}
				\STATE Use the constrained k-means method to cluster $\{y^*_i\}_{i=1}^n$ into $L$ groups subject to the constraint
				that the size of any group is no smaller than $n_{\min}$;
		\STATE Cluster $\{\theta_i\}_{i=1}^n$ into $L$ groups $\Theta_1,...,\Theta_k$ according to the clustering results of $\{y^*_i\}_{i=1}^n$;
		%\STATE $L = L+1$;
		%\ENDWHILE
		%\STATE $L = l$;
		\STATE $\widetilde{\Theta} = \{\Theta_l\}_{l=l}^L$;
				\FOR{$l=1$ to $L$}
		\STATE $N(l) = \mbox{NumOf}(\Theta_l)$;
		\ENDFOR
		%\RETURN $\Theta,N$
	\end{algorithmic}
\end{algorithm}

%It is worth to note that using a large number of mixture components of the prior distribution could yield a more exact lower bound approximation but require more samples for entropy estimation. Thus, a trade-off between the number of prior partitions and the number of samples distributed to each components should be considered. In addition, appropriate method classifying samples from parameter space can be chosen according to the specific model and the distribution of observations. In the following numerical examples, however, k-means method has been applied to cluster samples from observed space, and hence the corresponding parameters that generate them. 

\section{Numerical Examples} \label{sec:examples}
In this section, we demonstrate 
the performance of the proposed design method with three  examples. In all these examples, comparisons are made between the proposed method and the D-posterior precision method (with the posterior samples generated via ABC).
The implementation details such as the algorithm parameters are provided in the supplemental material. 
%{\color{blue}We note that,} the main purpose of all these examples is to compare the performance of the ED methods, and in this respect information %regarding the implementation details and the computational cost (e.g., the number of samples used)
% is omitted due to the limited space. 
We also note that, the main purpose of the examples is to compare 
the performance of the ED methods, and detailed comparisons in terms of both performance and  computational cost will be reported elsewhere.

\subsection{A mathematical example}
To illustrate the limitation of the ABC based method, we first consider a toy problem with strongly 
non-Gaussian posterior. 
Specifically, consider the following generative model
$$y=G(\theta,d)(1+\epsilon_1)+\epsilon_2,\quad G(\theta,d)=\frac{1}{B(2,d)}\theta(1-\theta)^{d-1},$$
where $B(\cdot,\cdot)$ is the beta function, $\epsilon_1 \sim N(0,0.05^2)$ and $\epsilon_2 \sim N(0,0.05^2)$. 
In this example, the likelihood is actually available, 
$$p(y|\theta,d)= N(G(\theta,d),0.05^2(1+G^2(\theta,d))).$$
The prior is assumed to be uniformly distributed on the interval $[0,1]$. The design variable is chosen from $[2,100]$. 
The main purpose of this example is two-fold:
first since the likelihood is available in this problem, we can accurately evaluate {the} expected utility, and validate our approximation against it; second, the posterior distribution of this problem is strongly non-Gaussian,
and we thus can show that the KLD and the ABC methods yield very different design in this case. 
%Thus, the nested Monte Carlo can be applied to give accurate estimates of KLD expected utility function, with which we can demonstrate the effectiveness of our approaches. We also calculate the Kraskov's mutual information estimation as a comparison. 

\begin{figure}[h]
	\centering
%	\subfigure[]
{\includegraphics[width=0.85\linewidth]{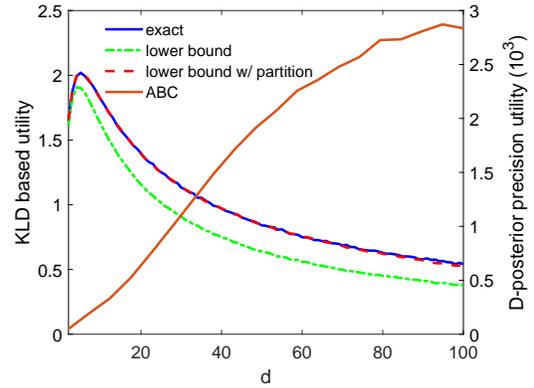}}\label{f:ex1kld}
%	\subfigure[]{\includegraphics[width=0.7\linewidth]{multi_abc.eps}}\label{f:ex1abc}
	%\subfigure[]{\includegraphics[width=4cm]{multi_d1.jpg}\includegraphics[width=4cm]{multi_d2.jpg}}
	\caption{The KLD expected utility plotted against the design parameter $d$,
	 compared to that based on the D-posterior, computed with ABC.} \label{f:ex1utility}
\end{figure}
\begin{figure}[h]
	\centering
	%\subfigure[]{\includegraphics[width=0.45\linewidth]{multi_d1.eps}}
	\subfigure[]{\includegraphics[width=0.72\linewidth]{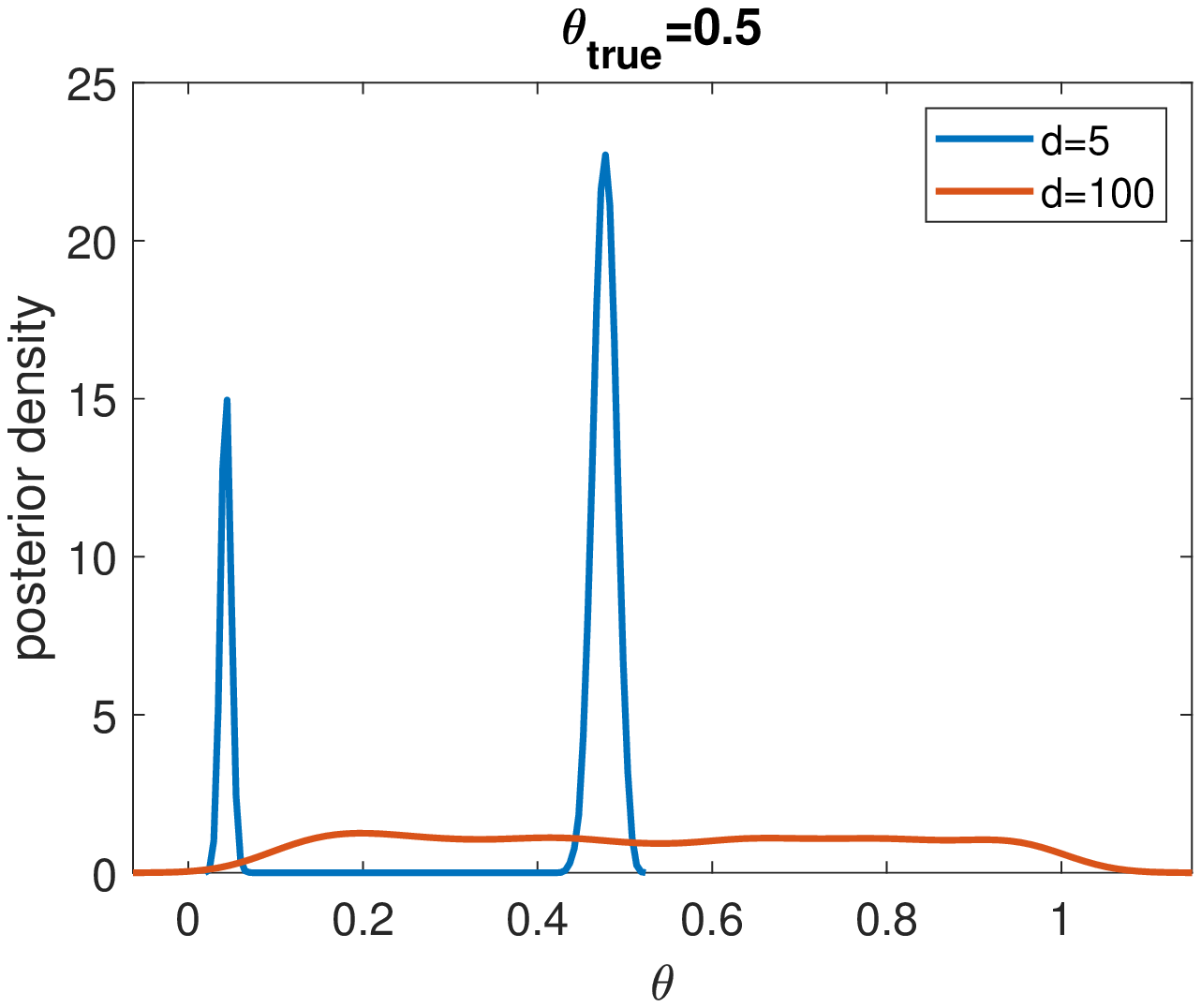}}
	\subfigure[]{\includegraphics[width=0.72\linewidth]{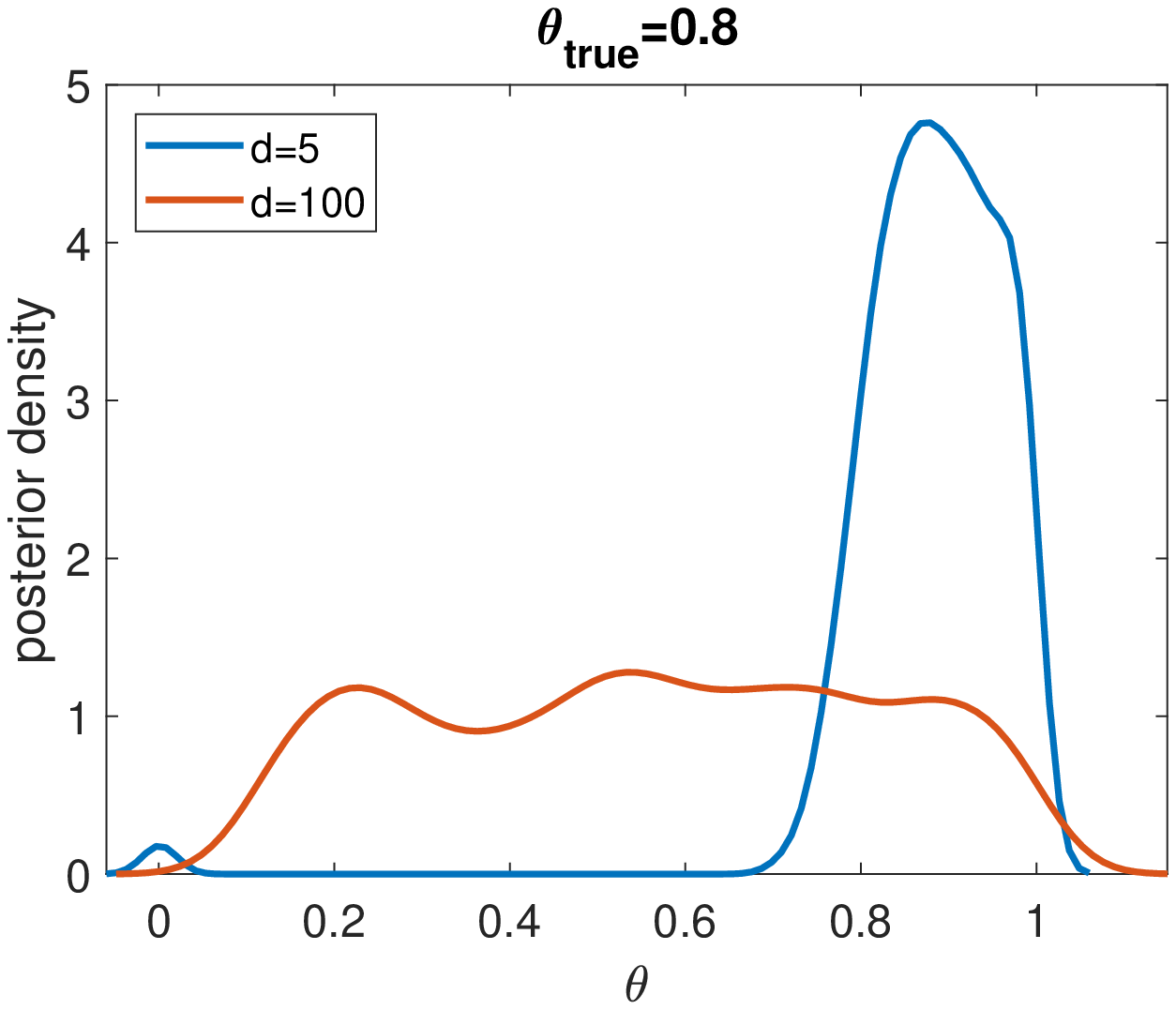}}
	%\subfigure[]{\includegraphics[width=4cm]{multi_d1.jpg}\includegraphics[width=4cm]{multi_d2.jpg}}
	\caption{The posterior distributions for $\theta_{true}=0.5$ (a) and 
	$\theta_{true}=0.8$ (b), obtained under the two experimental conditions $d=5$ and $d=100$.} \label{f:ex1posterior}
\end{figure}
First we estimate the KLD based expected utility function with the nested MC method, which is regarded as the exact value of the KLD.  
We then compute the lower bound approximations of the expected utility with and without the partition refinement.
We plot the expected utility computed by all the three methods as a function of the design parameter $d$ in Fig.~\ref{f:ex1utility}.
As one can see here that, the approximation computed with partition agree very well with the exact value 
of the expected utility, while that without partition admits obvious discrepancy from the exact value of the expected utility. 
Nevertheless, we can see from the figure that the optimal solutions predicted by all threes methods are  
largely the same $d=5$. As a comparison, in Fig.~\ref{f:ex1utility} we plot the expected D-posterior precision utility
as a function of $d$. 
Interestingly, in this case the expected utility is roughly an increasing function of $d$, and as a result 
the optimal solution is achieved at $d=100$, which is the upper bound of $d$.

To further compare the two methods, we conduct numerical experiments  
under the two experimental conditions: $d=5$ and $d=100$. 
We generate data and perform the Bayesian inference  for two cases: the true value is $\theta=0.5$ and 
the true value is $\theta=0.8$. We show the obtained posterior distributions in Fig.~\ref{f:ex1posterior}.
One can see here that the posterior distributions in this problem are greatly different from Gaussian 
and those obtained under $d=5$ are clearly bimodal with one mode concentrated around the ground truth. 
On the other hand, clearly the posteriors obtained under the condition $d=100$ look substantially less informative
than those under $d=5$, suggesting that the KLD utility is more effective for problems with strongly non-Gaussian posteriors.

\subsection{Ricker Model} 
Our second example is a ecological model describing the evolution of population size over time. The model assumes that the unobservable population size $N_t$ follows the scaled Ricker map \cite{turchin2003complex}:
$$N_{t+1}=rN_te^{-N_t+e_t},\quad t=1,...,T$$
where $e_t\sim N(0,\sigma^2)$ are independent process noise and r is a parameter related to the growth rate. The observation $Y_t$ is Poisson distributed as:
$$Y_t\sim Poisson(\phi N_t),$$
where $\phi$ is a scale parameter. We assume $\theta=(\log{r},\phi,\sigma)$ are the three parameters to be inferred and {the}
number of  observations is $T=50$. 
Following \cite{dutta2016likelihood}, the prior distribution is given by 
$$\log{r} \sim U(3,5), \,\phi \sim U(5,15),\,\sigma \sim U(0,0.6).$$
For convenience’s sake, one does not use the observations $y_1,...,y_t$ directly, and instead 
{we} select a number of summary statistics of the observations and use them as the data for the inference.  
Following \cite{wood2010statistical}, we set 13 summary statistics of $\{y_1,...,y_T\}$:  the average population and number of zeros observed over the given time, the autocovariances from lag0 to lag5, the coefficients $\alpha_0$, $\alpha_1$ and $\alpha_2$ of the quadratic regression $y_{t+1}=\alpha_2(y_{t+1}-y_t)^2+\alpha_1(y_{t+1}-y_t)+\alpha_0+\epsilon_t$, and the autoregression coefficients $\beta_0$ and $\beta_1$ based on the regression $y_{t+1}^{0.3}=\beta_0y_t^{0.3}+\beta_1y_t^{0.6}+\epsilon_t$. 
 For convenience's sake, we index these statistics as is in Table~\ref{tb:ex2}. The goal here
is to find a combination of two statistics in the 13 which can provide most information.
The likelihood function here is obviously not available in a closed form, and it is an often used example for likelihood-free inference~\cite{price2018bayesian,fearnhead2012constructing}. 

\begin{table}[h]
	\caption{numbered statistics}\label{tb:ex2}
	\vspace{.1in}
	\centering
	\begin{tabular*}{8cm}{c|c|c|c}
		No. & statistics&No. &statistics\\  
		\hline
		1& average population&	8& autocov lag5\\
		2& zeros observed& 9& $\alpha_0$\\
		3& autocov lag0& 10& $\alpha_1$\\
		4& autocov lag1&11& $\alpha_2$\\
		5& autocov lag2&12& $\beta_0$\\
		6& autocov lag3&13& $\beta_1$\\
		7& autocov lag4& & \\
	\end{tabular*}
	\vspace{.1in}
\end{table}

\begin{figure}[h]
	\centering
%	\subfigure[]{\includegraphics[width=4cm]{ricker_lb_1d.jpg}\includegraphics[width=4cm]{ricker_abc_1d.jpg}}
	\subfigure[]{\centerline{\includegraphics[width=.82\linewidth]{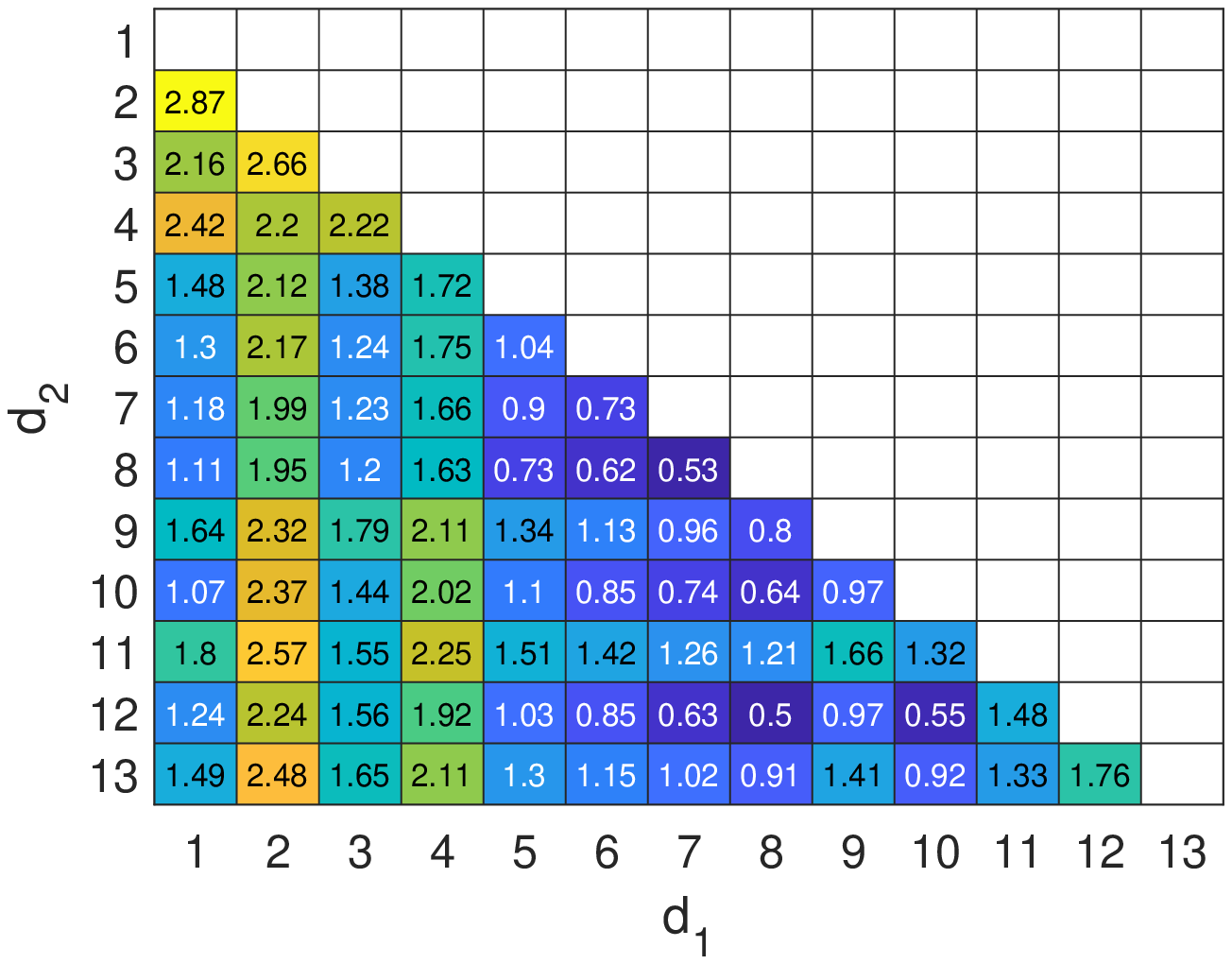}}}
	\subfigure[]{\centerline{\includegraphics[width=.82\linewidth]{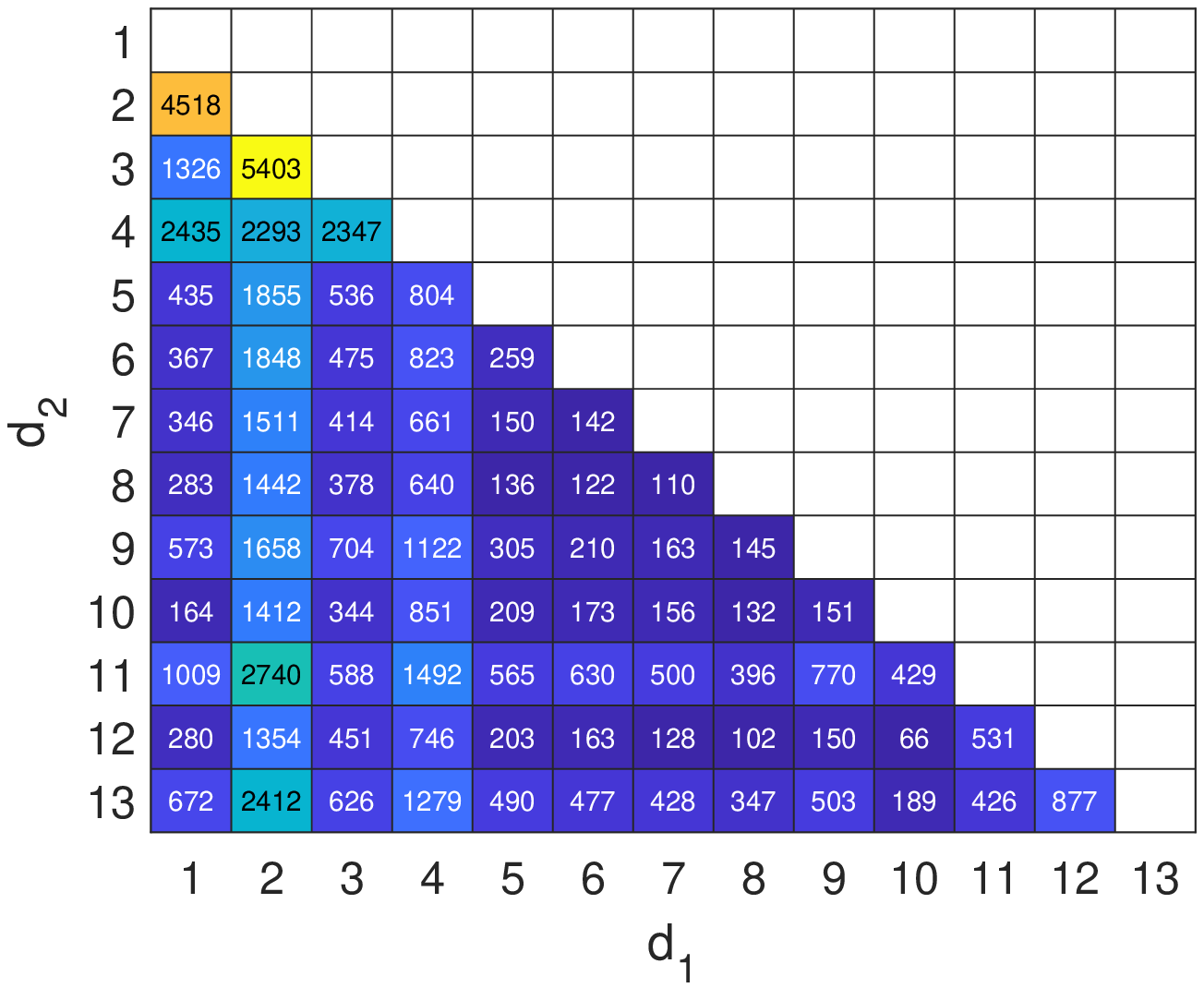}}}
	
	\caption{(a): the expected LB-KLD utility at all the scenarios.
	(b): the expected D-posterior precision utility at all the scenarios.}\label{f:ricker_utility}
\end{figure}

\begin{figure}[h]
	\centering
	\subfigure[]{\includegraphics[width=.9\linewidth]{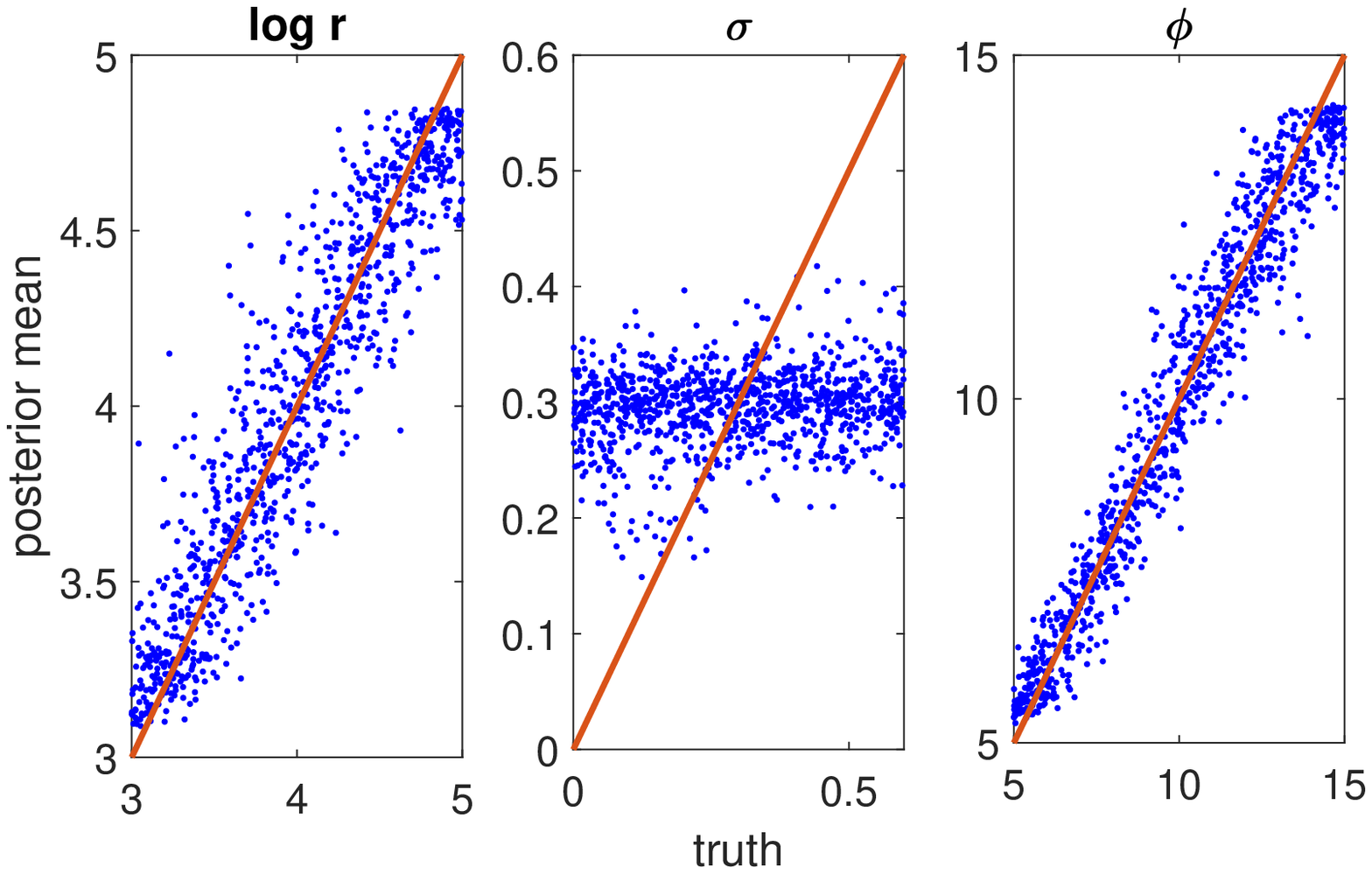}}
	\subfigure[]{\includegraphics[width=.9\linewidth]{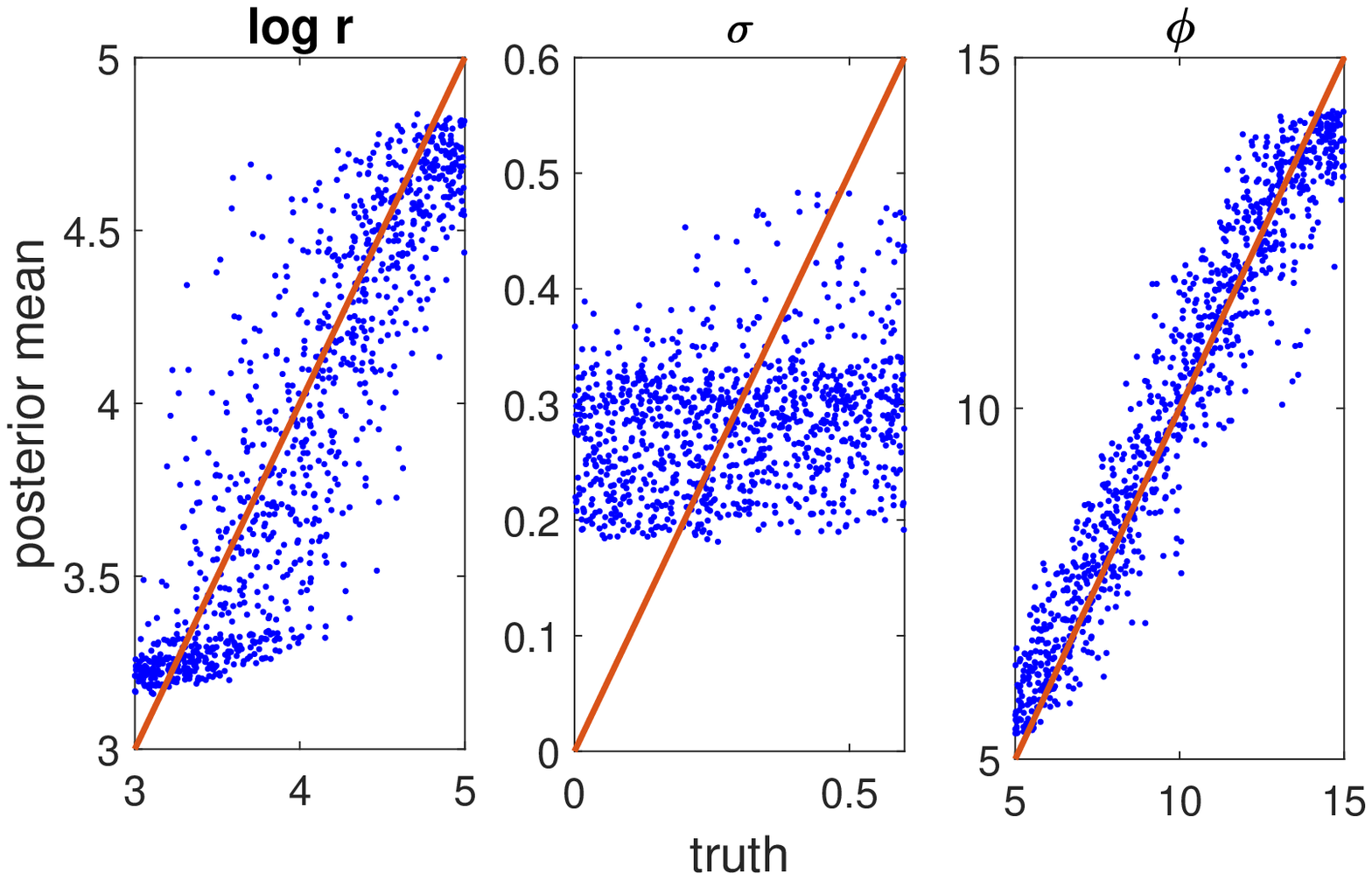}}
	\caption{The scatter plot of the posterior means against the true values for $d=(1,2)$ (the top) and $d=(2,3)$ (the bottom).}\label{f:ricker_post}
\end{figure}

We compute the expected LB-KLD utility at all the scenarios of the combination and show the results
in Fig.~\ref{f:ricker_utility} (a), 
which indicate that the optimal combination should be that of statistics (1) and (2).  
As a comparison, we also compute the expected D-posterior precision utility, also for all scenarios,
shown in Fig.~\ref{f:ricker_utility} (b), and with this utility function we 
obtain a different optimal combination, i.e., statistics (2) and (3). 
Thus, in this example, the two methods also result in different experimental conditions.  
To compare the performance of the two combinations, we conduct the following experiments. 
First we randomly generate 1000 pairs of truth $\theta_{true}$ and simulated data $y$ from the distribution $p(y,\theta)$.
With each pair of $(\theta_{truth},y)$, we conduct a Bayesian inference and compute the posterior distribution
using statistics (1) and (2), and using (2) and (3) respectively.
Since the likelihood is not available, all the posteriors are computed with the ABC method. 
In Fig~\ref{f:ricker_post} we show the scatter plots of the posterior means versus the true values of 
the 1000 trials for all the three parameters. 
Fig.~\ref{f:ricker_post} (a) shows the results from statistics 1\&2, which is the optimal condition determined by the LB-KLD method, 
and Fig.~\ref{f:ricker_post} (b) shows those from statistics 2\&3, the optimal solution by the D-posterior precision method.
The solid lines which represents that the posterior value is equal to the truth provides a guideline of the inference results,
and intuitively speaking, the points being distributed closer to the line indicates better inference results.   
In this respect, we can see that parameter $\sigma$ is much more difficult to infer than the other two parameters, in both designs. However, the figures show that design obtained by the LB-KLD utility seems to lead to better inference
results for $\log r$ and $\phi$. 
The advantage of the proposed method can also be quantitatively demonstrated 
by calculating the mean square error of the posterior means in the 1000 trials. While the MSEs for $\sigma$ are similar:
$0.080$ for LB-KLD and  0.079 for D-posterior precision, 
the design obtained by our method results in much lower MSEs for $\log r$ and $\phi$: 0.01 (LB-KLD) versus 0.02 (D-posterior) for 
$\log r$ and 0.0046 (LB-KLD) versus 0.0073 (D-posterior) for $\phi$ respectively. 

\subsection{Aphid Model}
The last problem we consider is a stochastic model describing the growth of aphid population~\cite{matis2007stochastic}. 
The purpose of this example is to illustrate that when the posterior distribution is not too far from Gaussian, the proposed method identify similar designs as the D-posterior precision approach. 
Let $N(t)$ and $C(t)$ denote the current live size and the accumulative size of aphid population. The population is assumed to grow with a rate of $\lambda N(t)$ and die with a rate of $\mu N(t)C(t)$ at any time $t$. Therefore, the probability that a birth or a death occurs in the next infinitesimal time period $\Delta_t$ is  
\begin{multline}\notag
    Pr\{N(t+\Delta_t)\!=\!n+1,C(t+\Delta_t)\!=\!c+1|N(t)\!=\!n,C(t)\!=\!c\}\\\!=\!\lambda n\Delta_t+o(\Delta_t),\\
\notag
	Pr\{N(t+\Delta_t)\!=\!n-1,C(t+\Delta_t)\!=\!c|N(t)\!=\!n,C(t)\!=\!c\}\\\!=\!\mu nc\Delta_t+o(\Delta_t),
\end{multline}
where $\lambda$ and $\mu$ are the birth and death parameters to be determined. The prior distributions for $\lambda$ and $\mu$ follow \cite{gillespie2019efficient}, where
$$
\left(
\begin{matrix}
\lambda \\ \mu
\end{matrix}
\right)
=N
\left[
\left(
\begin{matrix}
0.246\\0.000136
\end{matrix}
\right),
\left(
\begin{matrix}
0.0079^2& 5.8\times 10^-8\\
5.8\times 10^-8 & 0.00002^2
\end{matrix}
\right)
\right].
$$
And the initial aphid level is set as $N(0)=C(0)=28$. The goal here is to specify the sampling times $D=(t_1,...,t_k)$ so as to accurately estimate the model parameters.
This problem also does not admit a tractable likelihood function and so it can not be solved with the standard KLD based method.

\begin{figure}[h]
%	\vspace{.1in}
	\centering
	\subfigure[]{\includegraphics[width=0.65\linewidth]{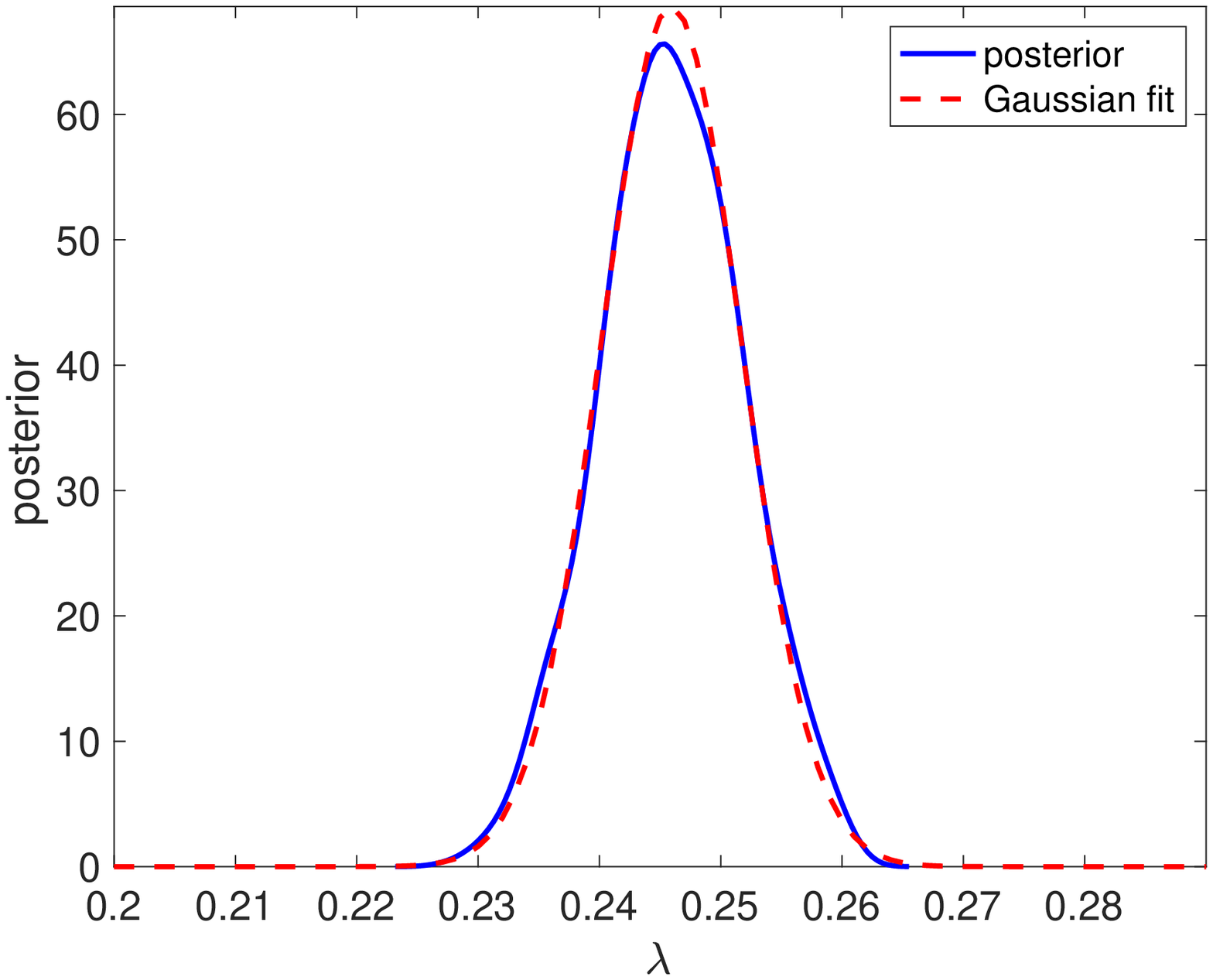}}
	\subfigure[]{\includegraphics[width=0.65\linewidth]{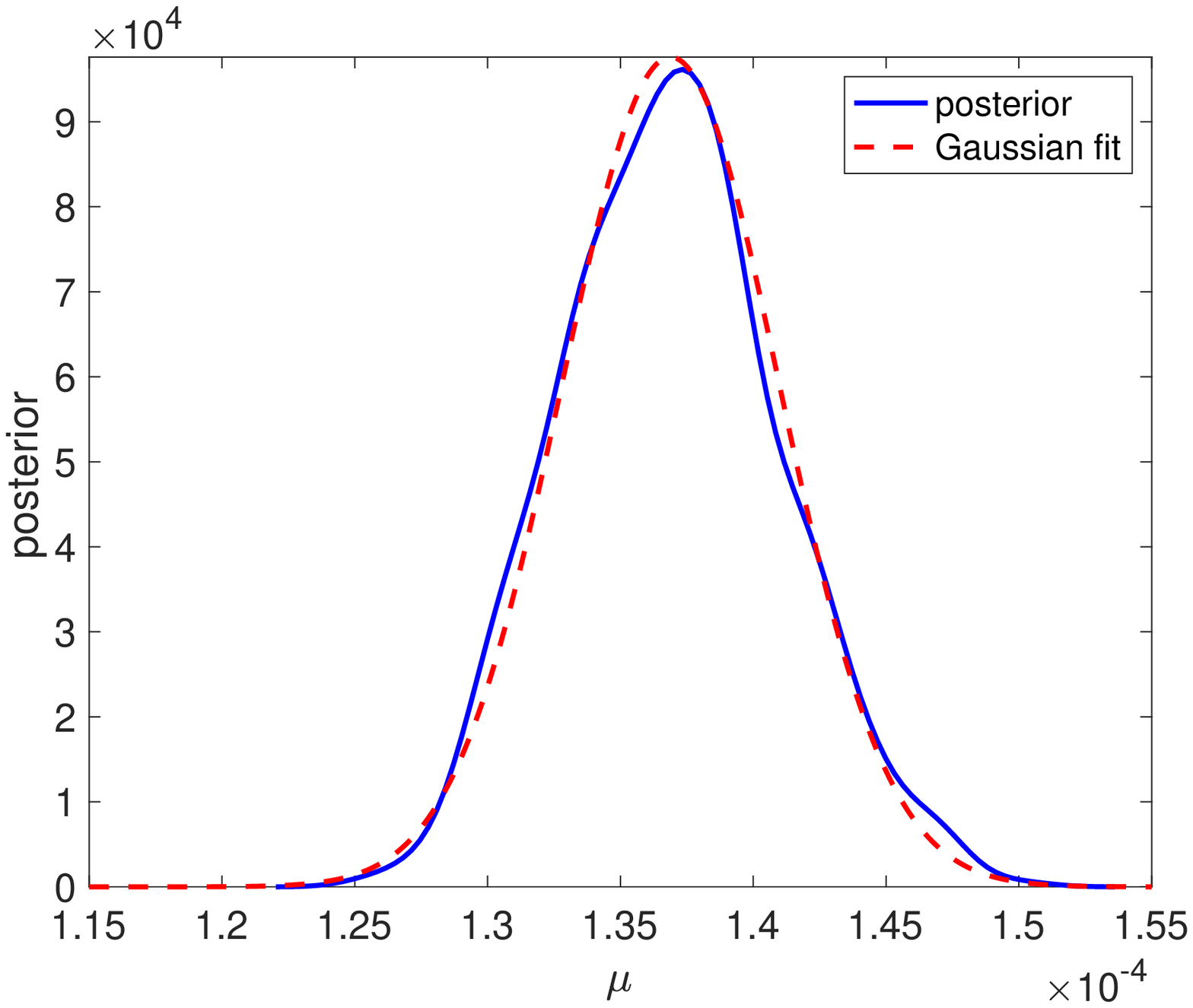}}
	\caption{(a): the posterior of $\lambda$ (solid) and its Gaussian fit (dashed).
	(b): the posterior of $\mu$ (solid) and its Gaussian fit (dashed).} \label{f:ex3post}
\end{figure}

We consider the observation times design in time interval $[0,50]$ which is equally discretized into 5000 grids
for the purpose of searching for the optimal designs. 
We then compute the optimal designs for $k=1,2,3,4$ using the LB-KLD and the D-posterior precision methods,
and the optimal designs are provided in Table~\ref{tb:ex3}
%{\color{blue}The optimal solutions are obtained by exhausting all the grid points.} 
One can see from the table that in this example, the methods 
actually identify very close experimental conditions.
To further analyze the methods, we conduct a Bayesian inference for the situation where the truth is located at the mean of the prior,
and the simulated data is collected at 4 time locations: $( 13.8, 19.1, 24.5, 30.6)$, which are the optimal choice selected by the LB-KLD method. 
We compute the posterior with ABC and plot the posterior distributions in Fig.~\ref{f:ex3post},
and also shown in the figure are the Gaussian fits of the posterior distributions. 
The figures show that the posterior distributions in this problem are very close to Gaussian, 
which provides a good explanation that the optimal designs obtained by both methods are rather close to each other.  

\begin{table}[h]
\caption{The optimal design results for the Aphid Model.} \label{tb:ex3}
	\vspace{.1in}
 \centering
  \small
 \begin{tabular*}{8.3cm}{c|c|c}
    \hline
    method & LB-KLD  & D-posterior \\  
    \hline
    1d& $(21)$& $(21)$\\
    \hline
    2d & $(17, 28$)& $(18, 27$)\\
    \hline
    3d  & $(15.7, 22.7, 32.0)$ & $(16.8, 21.9, 29.1)$ \\  
    \hline
    4d  & $( 13.8, 19.1, 24.5, 30.6)$ & $(15.8, 20.4, 25.2, 30.5)$ \\  
    \hline 
 \end{tabular*} 
\end{table}

\section{Discussion}
In this paper we have presented a Bayesian experimental design method for stochastic models with intractable likelihoods, and the method can be applied 
to problems where the posteriors are strongly {non-Gaussian}.  
Specifically we propose a new utility function, which is a lower bound approximation to the often used KLD utility, 
and we provide a entropy estimation based method to estimate the expected utility. 
Using numerical examples, we demonstrate that, the method performs well 
regardless whether the posterior is close to Gaussian, while the D-posterior precision method yields less effective designs when the posteriors differs strongly from Gaussian. 
We believe the method can be found useful in a large range of real world BED problems where the likelihood functions are intractable.

A number of improvements and extensions of the proposed LB-KLD method are possible. 
First as is mentioned earlier, in this work we only compare the performance against the D-posterior method, and a more comprehensive  comparison 
with many other aforementioned methods e.g. \cite{pmlr-v89-kleinegesse19a},
in terms of both effectiveness and efficiency, will be conducted   in a future work. 
Secondly, as the LB-KLD is essentially a lower bound approximation of the KLD utility, we anticipate that in certain circumstances, the method may not yield the same results
as the KLD, in which case, the approximation method is deemed unsuitable. 
Thus in the future we hope to understand the applicability as well is the limitations of the method. 
Last but not least, the prior partition procedure is an essential component of the proposed method.
In this work we provide an empirical partition strategy based on clustering the output samples, and an important  question here is that what is the optimal partition strategy (including the number of partitions 
$L$)  in theory?  
%A related theoretical question is that if one can derive an error bound between the LB-KLD (with partition) and the actual KLD utilities?  
%Also, is it possible to adaptively refine the partition to improve the estimation accuracy? 
We plan to investigate these issues in future studies.

\appendix
\section*{Supplementary Materials}
\input{supp_final.tex}

\subsubsection*{Acknowledgements}
The work was partially supported by NSFC under grant number 11301337.

\bibliographystyle{apalike}
\bibliography{aistats}
%\begin{thebibliography}{}
%\setlength{\itemindent}{-\leftmargin}
%\makeatletter\renewcommand{\@biblabel}[1]{}\makeatother
%\bibitem{} J.~Alspector, B.~Gupta, and R.~B.~Allen (1989).
%    \newblock Performance of a stochastic learning microchip.
%    \newblock In D. S. Touretzky (ed.),
%    \textit{Advances in Neural Information Processing Systems 1}, 748--760.
%    San Mateo, Calif.: Morgan Kaufmann.
%
%\bibitem{} F.~Rosenblatt (1962).
%    \newblock \textit{Principles of Neurodynamics.}
%    \newblock Washington, D.C.: Spartan Books.
%
%\bibitem{} G.~Tesauro (1989).
%    \newblock Neurogammon wins computer Olympiad.
%    \newblock \textit{Neural Computation} \textbf{1}(3):321--323.
%\end{thebibliography}

\end{document}

%% file: supp_final.tex
%\centerline{SUPPLEMENTARY MATERIAL}

\section{Proof of Theorem 3.1}
%\newtheorem*{thm}{\bf Theorem 3.1}
%\begin{thm}\label{thm1}
%	Suppose that  $\theta$ is a random variable defined on state space $\Theta$, with probability density $p(\theta)$. For any given $\theta\in\Theta$, let $y$ and $y'$ be two random variables that are independent conditional on $\theta$, and 
%	both follow the same distribution $p(y|\theta)$. Now define $z=y-y'$, and we then have,  
%	$$E_\theta[H(p(y|\theta))] \leq H(E_{\theta}[p(z|\theta)])-\frac{dim(y)}{2}\log2,$$
%	where $dim(y)$ is the dimensionality of $y$.
%	%Moreover, equality holds if and only if $X|\theta$s are Gaussian with the same probability density for different $\theta$s.
%\end{thm} 

%and the concavity of entropy.
%\newtheorem{Lemma}{\bf Lemma}
%\begin{Lemma}[entropy power inequility]\label{lm1}
%	 Suppose $y$ and $y'$ are independent p-dimensional real-valued random variables and $H(y)$ is the differential entropy of the probability density function $f_Y$
%	$$H(y)=-\int f_Y(y)\log[f_Y(y)]dy.$$
%	 Then
%	$$\exp(2H(y+y')/p)\geq \exp(2H(y)/p)+\exp(2H(y')/p).$$
%\end{Lemma}

%Define the probability density function of $y$ conditional on $\theta$ as $f(y|\theta)$ and that of $z$ as $g(z|\theta)$. Then the one of $-y'$ is $f(-y|\theta)$. Using Lemma~\ref{lm1} we obtain that,
\begin{proof}
	From  Shannon's entropy power inequality~\cite{cover2012elements}, we obtain, 
	\begin{equation*}
		\begin{aligned}
			&\exp(2H(p(z|\theta))/dim(y))\\
			\geq  &\exp(2H(p(y|\theta))/dim(y))+\exp(2H(p(-y|\theta))/dim(y))\\
			=  &2\exp(2H(p(y|\theta))/dim(y)),
		\end{aligned}
	\end{equation*}
	which implies that 
	\begin{equation}
	H(p(y|\theta))\leq H(p(z|\theta))-\frac{dim(y)}{2}\log2.\label{e:ineq1}
	\end{equation}
	
	Taking expectation with respect to $p(\theta)$ on both sides of Eq.~\eqref{e:ineq1} yields, 
	\begin{equation}
	\begin{aligned}
	&E_\theta[H(p(y|\theta))]\\
	\leq& E_\theta[H(p(z|\theta))]-\frac{dim(y)}{2}log2\\
	\leq& H(E_{\theta}[g(z|\theta)])-\frac{dim(y)}{2}log2,
	\end{aligned}
	\end{equation}
	where the last inequality is due to the concavity of the entropy~\cite{cover2012elements}. 
\end{proof}

\section{Proof of Corollary 3.2}
%\newtheorem*{Corollary}{\bf Corollary 3.2}
%\begin{Corollary}
%		Suppose $p(\theta)$, $p(y|\theta)$, and $p(z|\theta)$ are defined as is in Theorem~3.1, and $p(\theta)$  admits the form of,  
%		$$p(\theta)=\sum_{l=1}^{L}\omega_l f_l(\theta),$$
%		where $\omega_l\geq0$ for $l=1...L$, $\sum_{l=1}^{L}\omega_l=1$, and $f_l(\theta)$ are density functions.  
%		Then 
%		\begin{align*}
%		E_\theta[H(p(y|\theta))] &\leq \sum_{l=1}^{L} \omega_l H(E_{\theta\sim f_l}[p(z|\theta)])-\frac{dim(y)}{2}\log2\notag\\
%		&\leq H(E_{\theta}[p(z|\theta)])-\frac{dim(y)}{2}\log2.
%		\end{align*}
%\end{Corollary}
\begin{proof}
	Recall that the prior takes the form of $$p(\theta)=\sum_{l=1}^{L}\omega_l f_l(\theta),$$
	and we have
	\begin{equation}
	\begin{aligned}
	E_\theta[H(p(y|\theta))] =&\int_{\Theta}p(\theta)H(p(y|\theta)) d\theta\\
	=&\sum_{l=1}^{L} \omega_l \int_{\Theta} f_l(\theta) H(p(y|\theta)) d\theta\\
	\leq& \sum_{l=1}^{L} \omega_l H(E_{\theta\sim f_l}[p(z|\theta)])-\frac{dim(y)}{2}log2,
	\end{aligned}
	\end{equation}
	where the inequality above is a direct consequence of Theorem 3.1. 
	Once again, because the entropy is concave, we have
	\begin{equation}
	\begin{aligned}
	&\sum_{l=1}^{L} \omega_l H(E_{\theta\sim f_l}[p(z|\theta)])-\frac{dim(y)}{2}log2\\
	\leq & H(\sum_{l=1}^{L} \omega_l E_{\theta\sim f_l}[p(z|\theta)])-\frac{dim(y)}{2}log2\\
	=&H(E_{\theta}[p(z|\theta)])-\frac{dim(y)}{2}log2.
	\end{aligned}
	\end{equation}
\end{proof}

\section{Implementation details}
%We demonstrate the proposed method in three cases: (a) a mathematical example, (b) the Ricker Model, and (c) the Aphid Model. The first model is fast to simulate, so the numerical utility values are averaged over many repeated trials when comparing to the results obtained from nested MC method, which are intended as baselines. The last two models are relatively slow to simulate and likelihood-intractable. Thus, we estimate each utility value with a single run, and only empirically compare the performances (it is not possible to apply the nested MC method in these cases). 
This section provides the experimental setup and implementation details of the examples.
Code for reproducing our experiments can be found at \url{https://github.com/ziq-ao/LBKLD_estimator}.

{\bf The mathematical example.} \quad We estimate the expected LB-KLD utility function values with $3\times 10^4$ (i.e. $n=10^4$) model simulations. In the prior partition step, we set $n_{min}=10$ and $L=5$.  Averaging was done over 100 independent runs to mitigate the random errors. Moreover we generate a larger number ($10^5$) of samples to estimate the KLD based expected utility function values with the nested MC method. For the D-posterior precision method, 100 samples are kept from $10^4$ prior-predictive simulations to form the ABC posterior. Again, the reported results are the average over 100 runs.

{\bf Ricker Model.} \quad We estimate the expected LB-KLD utility with $3\times 10^4$ model simulations. In the prior partition step, we set $n_{min}=50$ and $L=5$. For the D-posterior precision method, 100 out of $10^4$ prior-predictive samples are used to compute the posterior statistics. 

{\bf Aphid Model.} \quad The implementation setup of the LB-KLD and the D-posterior methods is the same as that 
of the Ricker model. 
It should also be mentioned here that, for $k=1$ and $k=2$, the optimal solutions are obtained by exhausting all the integer grid points, while the Simultaneous Perturbation Stochastic Approximation algorithm~~\cite{spall1998overview} is used to optimize the expected utility functions for $k=3$ and $k=4$.